\documentclass[prd,superscriptaddress, superscriptaddress,amsmath,twocolumn]{revtex4}

\usepackage{epsf,latexsym,bbm,bm}
\usepackage{amssymb,amsmath}
\usepackage{times,graphics}
\usepackage{color,cancel}
\usepackage{epsfig,ulem}
\usepackage{euscript}
\def\6{{\langle}}
\def\9{{\rangle}}

\def\eH{\EuScript{H}}

\newcommand{\tr}{\mathrm{tr}}

\def\sg{\textsl{g}}

\newcommand{\be}{\begin{equation}}
\newcommand{\ee}{\end{equation}}
\newcommand{\ba}{\begin{eqnarray}}
\newcommand{\ea}{\end{eqnarray}}

\begin{document}
\title{The information recovery problem}
\author{Valentina Baccetti}
\affiliation{Department of Physics \& Astronomy, Macquarie University, Sydney NSW 2109, Australia}
\author{ {Viqar Husain}}
\affiliation{Department of Mathematics \& Statistics, University of New Brunswick, Fredericton NB E3B5A3, Canada}
\author{Daniel R. Terno}
\affiliation{Department of Physics \& Astronomy, Macquarie University, Sydney NSW 2109, Australia}

\begin{abstract}
The  {issue of} unitary evolution during creation and evaporation of a black hole remains {controversial}. We~argue that  some prominent cures are more troubling than the disease, demonstrate that their central element---forming of the event horizon before the evaporation begins---is not necessarily true, and describe a fully coupled matter-gravity system which is manifestly unitary.
\end{abstract}
\maketitle

\section{Introduction: Unitarity Lost}
\label{sec1}

Black hole physics became a {particularly} fascinating area of study  with {the} discovery of Hawking radiation \cite{h:74}. {Its emission  completes a thermodynamic picture of black holes, but also leads to the infamous information loss problem \cite{h:76,bmps:95,fn:98,wald:01lrr, modern,modern2}. % an %initial state of matter collapses into  a black hole, which then evaporates into thermal radiation

 {Following \cite{wald:01lrr} } the problem can be stated as follows: A black hole evaporates (completely or to a Planck-scale remnant)  within a finite time. The initial state of the collapsing matter had a low entropy. If the correlations between the inside and outside of the black hole are not restored during the evaporation, then by the time it has terminated, an initial low-entropy state will have evolved into a high-entropy state, implying that some ``information'' will have been lost. % Another causality of the evaporation process is the baryon number non-conservation.

Over the years numerous information-theoretic considerations have been applied {to this problem}. {In the meantime
quantum information  {theory} became an established cross-disciplinary field \cite{qinfo,qinfo2}, and~its impact   ranges from  the first technological applications to changing ways we think about other areas, including gravity \cite{rqi,rqi2}. }

Typically, the information loss and recovery arguments employ quantum fields on a fixed background spacetime.  This approach excludes the back-reaction of radiation on  the  metric, while~allowing  the black hole mass  to become time-dependent. The matter Hilbert space is often represented as $\eH\equiv\eH_M\otimes\eH_{\mathrm{in}}\otimes\eH_{\mathrm{out}}$, where $\eH_M$ is the Hilbert space of the infalling matter, and $\eH_{\mathrm{in}}$ and $\eH_{\mathrm{out}}$ are the Hilbert spaces of the ingoing and outgoing Hawking radiation.

Figure~\ref{figure1} presents the  conjectured  ``information  loss''  Penrose diagram describing this process. It pictures a black hole  evaporation as a sequence of  Schwarzschild black holes that terminates in a thermal matter  state in flat spacetime.  That is a lot of assumption:    a spherically symmetric metric for which Figure~\ref{figure1} is the Penrose diagram  might be parameterized { using the null coordinates} as
\be
ds^2 = -g(u,v) dudv + r^2(u,v) d\Omega.
\ee

However, the locus of points $r(u,v)=0$ would have to be (i) regular for advanced time $v <  v_0$, for~some $v_0$; (ii) singular for $v_0 \le v \le v_1$; and (iii) regular again for $v>v_1$. Thus~without a mechanism for dynamical singularity avoidance, this diagram represents an unlikely and computationally unjustified {physical} scenario.

\begin{figure}[!ht]{}
 \centering
   \includegraphics[width=0.32 \textwidth]{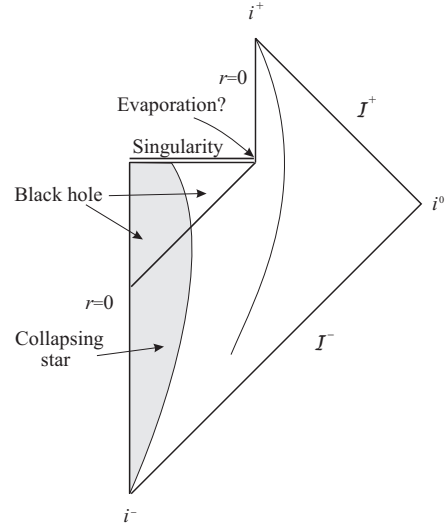}
     \caption{A conceptual description of black hole formation and  evaporation. Trajectory of an outside observer is marked as a thin line that terminates at the future timelike infinity $i^+$. } \label{figure1}\vspace{-6mm}
\end{figure}

These circumstances have not deterred  using Figure~\ref{figure1} as the starting point for   solving  the information loss problem. The goal is to come up with ways in which unitarity  can be rescued by extracting information from behind the horizon. (See \cite{modern,modern2} for recent reviews).

We argue in {Section \ref{sec2}} that some popular solutions are not only incomplete, but also raise  conceptual issues  that are potentially as serious as the original information loss problem.  Although a quantum theory of gravity may require dramatic changes of quantum theory,  {we think there is no need to postulate them as yet.   {In Section \ref{sec3}} we show   a simple model  where evaporation   prevents the horizon from even forming. While not making the entropy increase of matter any smaller, it makes clear that taking into account the full matter-gravity system is essential. {Section 4 presents a model of a coupled scalar field and gravity in spherical symmetry. The classical analysis of this system results in the Hamiltonian constraint and the Hamiltonian, making its quantum counterpart unitary by construction.} {Discussion of Section \ref{sec5}  points that} Figure~\ref{figure1} is the wrong starting point, {and outlines    new information-theoretical aspects to the entropy problem.}

We use the $(-+++)$ signature of the metric and set $c=\hbar=G=k_B=1$.

\vspace{-1mm}
%%%%%%%%%%%%%%%%%%%%%%%%%%%%%%%%%%%%%%%%%%
\section{Scrambling for Information}
\label{sec2}

Three key quantum features---no-cloning,  monogamy of {entanglement} \cite{qinfo,qinfo2,h4:09}, and a lesser-known no-disentanglement \cite{t:99}  results are often violated by  proposals of information recovery that begin with Figure~\ref{figure1}.
%Reference 11 is missing, please check and revise

The no-cloning theorem---that no  {unitary, or a more general completely positive  (see Section~\ref{sec5})} quantum process allows $\rho\rightarrow\rho\otimes\rho$---appears to be violated by postulating matter unitarity:  if~information  contained in the infalling matter turns up in the outgoing radiation, then  cloning of  information must have occurred. This  problem is apparently  resolved by the principle of observer complementarity \cite{s:93}; it is the statement that  unverifiable cloning behind a horizon is not an issue.

Another serious problem arises from considering entanglement monogamy: given two strongly entangled systems $\eH_A$ and $\eH_B$, neither can be  strongly entangled with a third system $\eH_C$.   \mbox{Almheiri {et al.} (AMPS) \cite{amps}} produced an observer that could witness strong entanglement between modes crossing the horizon  ($\eH_A$) and early emitted radiation ($\eH_B$) on the one hand, and between the early and late radiation ($\eH_C$) on the other. Hence observer complementarity does not preserve monogamy.  {Their cure is to introduce severe ``back reaction" in the form of a firewall, while ignoring its effect on the geometry.}

There are issues with this rescue attempt. One potential issue is that  while the firewall  saves monogamy, the cost is that it discards the equivalence principle \cite{bpz:13,kate}. Moreover, while entanglement is a  fragile quantum resource, perfect disentanglement, which takes an entangled state into the direct product of its reduced states
 \be
 \rho_{AB}\rightarrow\rho_A\otimes\rho_B, \qquad \rho_A=\tr_B \rho_{AB}, \quad \rho_B=\tr_A \rho_{AB}
 \ee
 cannot be realized by a linear quantum process \cite{t:99}.  {Among the byproducts of disentanglement are  violation of maximal bounds on our ability to distinguish non-orthogonal quantum states \cite{t:99,mt:00}}, cloning, and  violation of uncertainty relations \cite{mile:15}.   From the quantum-informational  view,  the firewall    is designed to be a universal disentangler  with its accompanying adverse consequences. Indeed, the firewall  in at least one concrete toy model \cite{jorma:14} behaves as disentangler.

Another idea is the ``$ER=EPR$''  proposal \cite{ms:13}, which postulates a wormhole connecting the interior of a black hole to the asymptotic region outside the horizon. This gives  information recovery  through  nontrivial topology. Among other issues this likely violates the topological censorship theorem~\cite{topc}.

Setting up a wormhole is a non-trivial task. Let us consider the constraint equations of general relativity. The simplest solution for scalar matter (phase space variables $\phi, P_\phi)$ is obtained by setting (ADM momentum) $\pi^{ab}=0$ and $P_\phi=0$. This immediately solves the spatial diffeomorphism constraint. Then with the Misner ansatz \cite{misner}
\be
ds^2 = \psi^4(r)\left( dr^2 + r^2 d\Omega^2\right),
\ee
the hamiltonian constraint becomes
\be
\nabla^2\psi + \frac{1}{8} (\phi,_r)^2 \psi =0.
\ee

With $\phi=0$, the simplest solution is the wormhole $\psi(r)= 1+ m/r$. With matter, an easy way to solve this equation is to take a distorted wormhole function $\psi(r)$ and find $\phi$.  This is possible only if $\nabla^2\psi$ is negative or zero for $r\in (0, \infty)$, which never happens: the function $r^2\psi^4(r)$ must be concave up and positive everywhere.  This means $\psi'>0$ and $\psi''>0$, hence $\nabla^2 \psi >0$. Thus there is no scalar field supporting a wormhole at a moment of time. Although the scalar field  is classical, appeal to the quantum violation of the energy conditions is impossible: localization of  quantum scalar particles can only be described using  the positivity of their energy densities \cite{local,local2}.

Yet another idea is the final state solution, proposed by Horowitz and {Maldacena} \cite{hm:04}, and  recently developed by Lloyd and Preskill \cite{lp:14}.  Here, information propagates with the collapsing matter from past infinity to the singularity inside the black hole, where it is scrambled and reflected, propagates backwards in time to the horizon, and then forward in time from the horizon to future infinity.

This picture becomes particularly elegant when {presented in terms of teleportation} using the  consistency framework for  closed time-like loops.  {Its } rules for probability calculations in presence of the closed time-like curves are based on the conditional probabilities with the prescribed measurement outcomes  (in the usual  teleportation procedure) \cite{l:ctc}. However, now, the price for information recovery includes non-linear evolution of  the chronology-preserving states (that may be unobservable), and~restrictions on the possible operations that the observer is able to perform beyond the horizon.

This connects with the $ER=EPR$ approach, which also appears to violate the monogamy of entanglement.
Constraining the allowed states and introducing certain identification between the spaces resolves this problem \cite{bv:14}. However, extending this operation to mixed states results in disentanglement. This is a grave violation if performed on a time slice. However, it is a natural feature of  time travel, thus supporting the argument of \cite{kate}.

 {A different group of arguments is based on the tunneling
picture of Hawking radiation and investigation of higher orders of the expansion in $\hbar$ of quantum corrections and backreaction \cite{pw:00,svzr:10,bp:11,zczy:11}. In particular, entanglement of the modes across the horizon is found to be necessary for preservation of both unitary evolution of black holes and the equivalence principle \cite{bpz:13}.  Aspects of intermode correlations and entanglement and their role in restoration of unitarity are reviewed in \cite{ijmpd:13}. }

%%%%%%%%%%%%%%%%%%%%%%%%%%%%%%%%%%%%%%%%%
\section{Horizon Unattained}
\label{sec3}

 {The information loss problem has two key elements: transformation of a low-entropic matter into a highly entropic Hawking radiation, and presence of  horizons that allow for information to leak out through them and become lost. We now demonstrate that accepting evaporation may lead to the absence of a horizon.}

 {An event horizon is a classical notion that is used in the quantum arguments (for example, to~prescribe the backward evolution of the modes in derivations of the Hawking radiation, or to justify a particular tensor structure of  model Hilbert spaces that are used to analyze the information problem). Hence~consistency of  our investigation of the logic of the information loss requires to treat   motion of the collapsing matter (semi-)classically.}

  {We note first that} according to a distant observer (Bob at the space-like infinity)  a classical collapse into a black hole takes an infinite amount of time.  {Then}, accepting the Hawking radiation and Page's formula \cite{modern,modern2} that describes the mass loss as
\be
\frac{dM}{dt}=-\frac{\gamma}{15,360\pi M^2}, \label{Page-evap}
\ee
where $\gamma$ is the number of discrete degrees of freedom of evaporating particles, means that the evaporation takes only a finite amount of Bob's time, $t_E$.  {(This simple expression is a late time result and does not include the grey-body factor \cite{wald:01lrr,modern,modern2}. It is not going to be used in the following.})

For the information loss problem to be meaningful, one of the following alternatives must be realized. One possibility is  that the  quantum effects are not strong enough to facilitate a finite-time collapse (i.e., crossing of a suitably defined horizon within a finite  time according to Bob). In this case, if~we accept that the Hawking radiation is observed by Bob, evaporation should start when the collapse brought the matter  suitably close to the Schwarzschild radius.

 On the other hand, it is possible that quantum effects are responsible for  crossing of the Schwarzschild radius $r_\sg=2M$ in the finite time. This (tautologically) implies that they are important in some neighborhood around the ``would-be'' horizon. Then it is not necessary to assume that any radiation is emitted before the horizon is crossed. However, this absence of radiation still would  imply  that the horizon region of even  a big black hole is  non-classical (either a test particle that is dropped in just after the collapsing matter will also cross (the fully formed) horizon, or the ``original'' collapsing matter  behaves differently from everything else that follows it).  {Since derivations of the Hawking radiation do not assume exotic classical background structures \cite{h:74,h:76,bmps:95,fn:98,wald:01lrr,modern,modern2}, the study of its consequences should not assume them either}.

 The arguments that the collapse according to Bob is never complete and overlaps with the onset of Hawking radiation that begins when the collapsing matter concentrates near
the gravitational radius were already made in \cite{ger:76}. {While the horizon is essential in the most intuitive derivations of the black hole radiation \cite{h:74, bmps:95, fn:98}, numerical analysis of  collapsing thick shells \cite{acvv:01},  {and a number of analytical investigations \cite{liberati:06, liberati:08, vsk:07, kmy:13}},  support emission of the pre-Hawking radiation}.

Hence we assume an early onset of   evaporation and investigate its consequences in a very simple system---a collapsing thin shell \cite{poisson}. %Classically it is simple enough to have an analytical solution.
We consider a spherically symmetric collapse of  a uniform thin shell $\Sigma$  {in 3 + 1 dimensional spacetime}, whose trajectory is parameterized by a comoving observer Alice. In the classical (non-evaporating) setting we can parameterize the shells' trajectory as $\big(t=T(\tau)$, \mbox{$r=R(\tau)\big)$} in the Schwarzschild coordinates, or as
$\big(u=U(\tau), r=R(\tau)\big)$ in the Eddington-Finkelstein coordinates, where
\be
u:=t-r+C\ln\left(\frac{r}{C}-1\right), \label{ef-s}
\ee
 {where $r_\sg=2M=C$.}
Inside the shell the spacetime is Minkowskian. This model is simple enough to have an analytic solution $\tau(R)$.

Our treatment of the quantum version of this model is based on the following assumptions:
\begin{enumerate} 
\item The classical spacetime structure is still meaningful and is described by a metric $g_{\mu\nu}$.
\item The classical concepts, such as trajectory, event horizon or singularity can be used.
\item The metric is modified by quantum effects. The resulting curvature satisfies the semiclassical~equation
\be
G_{\mu\nu}=8\pi \6\hat{T}_{\mu\nu}\9, \label{semiein}
\ee
where $G_{\mu\nu}$ is the Einstein tensor corresponding to the metric $g_{\mu\nu}$ and $\6\hat{T}_{\mu\nu}\9$ is the expectation value of the stress-energy tensor.
\item The collapse leads to a pre-Hawking radiation.
\end{enumerate}

{The first three items are the standard assumptions in the Hawking radiation and the related problem. Specifically, the diagram of Figure~\ref{figure1} makes sense only if the Assumptions 1 and 2 are valid. The last assumption summarizes the preceding discussion.}

{Modifications of a  thin shell collapse are then quite straightforward. Moreover, using this model allows to avoid the complications related to the place of origin \cite{giddings:16} of evaporating quanta. Unlike~previous works (e.g., \cite{dfu:76,ss:15}) that considered such models by focusing on the right hand side of Equation~\eqref{semiein}, we are primarily concerned with the effects on the shell's trajectory. We assume that the quantum effects are summarized as an appropriate evaporation law and incorporate this information into a metric that describes geometry outside the evaporating shell. The expectation of the stress energy tensor is then calculated via Equation~\eqref{semiein}.}

We model the spacetime outside via
 the outgoing Vaidya metric \cite{bmps:95,vai:51,fw:99}, a popular metric to describe a spacetime of radiating non-rotating center,
\be
ds^2=-f(u,r) du^2-2du dr+r^{2} d\Omega, \qquad f(u,r)=1- C(u)/{r}. \label{met2}
\ee
In the case of evaporating shell the relationship \eqref{ef-s} is no longer valid.

This is not the most general spherically-symmetric metric. A different metric that allows to explicitly  incorporate the evaporation law Equation~\eqref{Page-evap} and more general scenarios are considered in~\cite{bmt:16}.

 {We do not assume any specific form of the evaporation law, even if the linear evaporation $C=a-b u$ \cite{fw:99} is plausible}.  {As the equations below demonstrate} the effect of evaporation becomes appreciable when $R(\tau)=C+\epsilon$, for some small ($\epsilon\sim 1$) distance from the Schwarzschild radius, it is enough to assume that $C_u:=dC/du\leq 0$.  {A natural scale of the problem is set by $C$ and $C_u$}.

The junction conditions \cite{poisson} on metric and extrinsic curvature   lead to the equation of motion of the collapsing shell parameterized by the proper time $\tau$,
\begin{align}
&\frac{1}{8 \pi}\left(\frac{2\ddot R+F'}{2\sqrt{F+\dot R^2}} -\frac{\ddot R}{\sqrt{1+\dot R^2}}+  \frac{\sqrt{F+\dot R^2}-\sqrt{1+\dot R^2}}{R}\right. \nonumber \\
&\qquad\qquad\qquad\left. -F_U \dot{U} \left[ \frac{\dot R}{2F\sqrt{F+\dot R^2}} -\frac{1}{2F}\right]\right) = 0,
 \end{align}
where the last term is the result of evaporation. Here
\be
F=f(u,r)|_\Sigma=1-\frac{C(U)}{R}, \qquad \left.F_U=-\frac{1}{r}\frac{dC}{du}\right|_\Sigma,
\ee
and
\be
\frac{dU}{d\tau}=\frac{-\dot R+\sqrt{F+\dot R^2}}{F}.
\ee

We {note that}
\be
\frac{dC}{d\tau}=\frac{dC}{dU}\dot U\leq 0, \qquad F_U\geq 0.
\ee

The key quantity is the coordinate distance between the shell and its Schwarzschild radius,
\be
x(\tau):= R(\tau)-C(\tau).
\ee

Expanding $\ddot R$ in the inverse powers of $x$ and $C$ gives
\be
\ddot R = \frac{2 \dot R^2 \sqrt{1 + \dot R^2}}{ \big (\dot R + \sqrt{1 + \dot R^2}\big)}\frac{C}{x^2}\frac{d C}{d  U} + {\cal{O}}(x^{-1}),
\ee
thus describing the ever accelerating collapse as the shell approaches the Schwarzschild radius $C$. Hence (when $|\dot R|\gg 1 $)
\be
\ddot R \approx {4 \dot R^4}\frac{C}{ x^2}\frac{d C}{d U}.
\ee

At the same time, since close to the Schwarzschild radius we have
\be
F\approx \frac{ x}{C}, \qquad \dot U\approx -\frac{2\dot R}{F}\approx -\frac{2\dot R C}{x},
\ee
the distance $x$ evolves according to

\be
\dot x  = \dot R \left(1- \frac{2C}{x} \left|\frac{d C}{d U}\right| \right)  = |\dot R | \left(\frac{\epsilon_*(\tau)}{x}  -1\right),
\ee
with
\be
\epsilon_*(\tau)=-2C(\tau) \left.\frac{d C}{d  u}\right|_{u=U(\tau)}>0.
\ee

As a result  for $x< \epsilon_*$, we are guaranteed $\dot x>0$, hence  {stopping the approach to} the shrinking Schwarzschild radius.   {From the estimate $x\sim\epsilon_*$ we get
\be
\ddot R \approx \frac{ \dot R^4}{C C_U}\sim\propto -\dot R^4 C,
\ee
if we assume that $C_u\propto C^{-2}$, as in Equation~\eqref{Page-evap}. Thus in terms of Alice's proper time the rate of collapse and evaporation accelerate, giving the runaway solution. Nevertheless, she never crosses the Schwarzschild radius.  {From the moment of time $\tau_*$, $x(\tau_*)=\epsilon_*(\tau_*)$ it shrinks faster than the shell collapses}. {In our model this is true for any non-zero evaporation rate.} For Bob, however, the process is seen as very long lingering at effectively the Schwarzschild radius}.
 This behaviour persists in all spatial dimensions $D\geq 3$ and is true for at least some other types of the metric \cite{bmt:16}.
 
\section{Unitarity Regained}}
\label{sec4}

We have seen that some popular approaches to information recovery bring with them undesirable features, including non-linearity.  Typically, non-linear quantum evolution is an effective feature of open system dynamics, described using only the variables of the (open) system itself. Moreover, requiring consistency of the information loss setting brings models in which the event horizon  never forms. This suggests that  preserving the usual rules of quantum mechanics requires including gravitational degrees of freedom to obtain a closed system. {The lost information may be stored in the matter-gravity correlations \cite{kay:98a, kay:98b}}.

{Unitary evolution obtains in quantum theory for the quantized classical hamiltonian systems. To demonstrate the viability of the hypothesis of matter-gravity correlations we need to establish a Hamiltonian, and not just a Hamiltonian constraint that are the standard feature of the gravitational systems. We do it in} a simple, but physically rich matter-gravity system: the Einstein-scalar field theory in spherical symmetry. Fully nonlinear classical studies of this system reveal critical behaviour at the onset of gravitational collapse:  a black hole forms initially with infinitesimal mass and then grows by accreting scalar field \cite{choptuik}.

 The problem  can be precisely set up in a Hamiltonian framework.
In the Arnowitt-Deser-Misner (ADM) Hamiltonian formulation for general relativity the phase space of the model is defined by prescribing a
form of the gravitational phase space variables $q_{ab}$ and
${\pi}^{ab}$, together with fall-off conditions  for these
variables, and for the lapse and shift functions $N$ and $N^a$.  The
bulk  ADM \mbox{3 + 1} action for general relativity minimally coupled to a
massless scalar field is
\be
S = \frac{1}{16\pi}\int d^3x dt\left(  {\pi}^{ab}\dot{q}_{ab} +
{P}_\phi\dot{\phi} - N H - N^a H_a\right).
\label{act}
\ee

The pair $(\phi,P_\phi)$ are the scalar field canonical variables, and the Hamiltonian and spatial diffeomorphism constraints, $H$ and $H_a$, coupled to the lapse function $N$ and the shift vector $N^a$~\cite{poisson} take their standard form.
This action (together with the boundary terms, {see e.g.,} \cite{poisson, hw-flat}) is well-defined and determines the fall-off conditions on canonical variables.
  The reduction to spherical symmetry utilizes an  auxiliary
flat Euclidean metric $e_{ab}$ and unit radial normal  $s^a= x^a/r$, where  $r^2=e_{ab}x^ax^b$.
The parametrization is given by two geometric dynamical variables, $\Lambda(r,t)$ and $R(r,t)$, and their canonical conjugates.
Hence the spatial metric is
\be
dl^2 = \Lambda^2(r,t) dr^2 + R^2(r,t) d\Omega^2.
\ee
%The Painleve-Gullstrand  (PG) coordinates are those  where equal coordinate time slices are
%spatially flat \cite{hw-flat,eric}.

It is sufficient to use the partial gauge fixing
$\Lambda=1$  to obtain  non-singular coordinates at the horizon, which is
the feature of PG coordinates we desire.

The ADM $3+1$ action with a minimally coupled
scalar field  leads to the reduced action and the reduced Hamiltonian and  radial diffeomorphism constraints.
These   constraints are first class with an algebra that is similar to that of the full theory  \cite{hw-flat}.

The  gauge choice $\Lambda=1$, which corresponds to a step toward flat slice coordinates.   With this gauge condition, the Hamiltonian constraint is solved (strongly) for the conjugate momentum $P_\Lambda$ as a function of the phase space variables.  This gives
\be
P_\Lambda = P_RR + \sqrt{ (P_RR)^2 - X},
\label{PL}
\ee
where
\be
X =  16R^2 (2RR'' - 1 + R'^2) + 16R^2 H_\phi,
\ee
and allows to represent the radial diffeomorphism constraint as
\be
H_r = -P_\Lambda' + P_R R' + P_\phi \phi' \simeq 0, \label{consop}
\ee
with $P_\Lambda$ given by (\ref{PL}) above. We note that using this constraint  the square root in the latter equation
can be written as
\be
 \sqrt{ (P_RR)^2 - X} = \int_0^r \left(    P_R R' + P_\phi \phi'  \right) -P_RR
 \label{root-expr}
\ee
while
\be
 H_\phi = \frac{P_\phi^2}{2R^2} + \frac{R^2}{2}\ \phi'^2.
 \ee

The evolution equation for $\Lambda$ \cite{hw-flat} and the requirement that the gauge $\Lambda=1$ be preserved under it leads to fixing of the
lapse $N$ as a function of the shift $N^r$.
Finally, the reduced gravitational Hamiltonian~is
\begin{align}
H_{R}^G =&\int_0^\infty (N^r)'\left( R P_R + \sqrt{(P_RR)^2 - X}\right) dr\nonumber \\
&+\int_0^\infty N^r(P_RR' + P_\phi \phi')\ dr,
\label{Hred}
\end{align}
where the surface term in the reduced action has been written as bulk term and combined
with the remaining radial diffeomorphism constraint.

The gravity phase space variables are $(\Lambda, P_\Lambda)$  and the scalar field ones are $(\phi,P_\phi)$. These are subject to the Wheeler-DeWitt equation, the quantum analog of the Hamiltonian constraint.  The arena for quantization is a Hilbert space
${\cal H}_G \otimes {\cal H}_\phi$ of gravity and the scalar field.

While the actual quantization is very hard (and any practical calculations may be impossible), two key conceptual features are clear.  {First,} the Wheeler-deWitt  equation for this system is a manifestly matter-gravity entangling equation:   product states become  entangled when acted upon by the  constraint operator $H_r$ of Equation~\eqref{consop}. This is why in describing the information  loss problem we referred to the matter state as ``low-entropy'' and not ``pure''.  {Second},  since the combined matter-gravity system has a true Hamiltonian \eqref{Hred}, and not just the Hamiltonian constraint, the corresponding quantum evolution must be {unitary} \cite{ht:10}.

\vspace{5mm}

\section{Discussion}
\label{sec5}

 {We have seen that some of the popular cures for the loss of unitarity bring with them side effects that are, arguably, even more severe than the original problem. Entropy generation in a subsystem is typically a sign of an interaction with the environment, and the efficient formalism of completely positive (CP) evolution \cite{qinfo, qinfo2} exists to deal with such settings. Unitary evolution is just a particularly convenient example of this more general dynamics.   However, perfect cloning of unknown quantum states is forbidden in this framework. A more benign-looking possibility of a mere increase of distinguishability {(such as may result from disentanglement \cite{t:99})} indicates that the dynamics is of a fundamentally different type---a non-CP-evolution \cite{t:99,ctz:08}. {If this is the consequence of the proposal to recover information, then the result is even more serious violation of the quantum formalism.}

 There are more approaches that depend on existence of the horizon. We have a very simple but fully consistent model of collapse that is affected by   radiation. In this model, for almost entire evaporation time the shell stays very close to its Schwarzschild radius, but never crosses it. As a result, there are no trapped surfaces, no horizon and no singularity. The steady-state distance $\epsilon_*$ is in the trans-Planckian regime, but no more so than is used in the derivations of the Hawking radiation. The model  {neither} assumes the thermal character of the Hawking radiation,  nor any specific form of the pre-Hawking \cite{vsk:07} radiation, but only its existence and observability in finite time. In addition to possible problems that the matter-only resolutions of information loss problem entail, those that rely on the existence of an event horizon become untenable.

A semiclassical analysis of Schwarzschild black holes indicates that quantum fluctuations effectively destroy the horizon \cite{brustein:14}. Black hole radiation is  directly linked to quantum fluctuations \cite{bmps:95,fn:98}. The model  of   quantum matter on a classical background spacetime with a sharply defined horizon is a useful idealization, but it does not correspond to the asymptotic (semi-classical) future of a collapsing matter. The model of Section~4 demonstrates this explicitly by having  $\epsilon_*>0$ at all times.   Absence of the horizon can be seen as demoting the information loss problem from being a paradox  to a (still intricate) calculational problem.  Instead of a logical contradiction that follows from combining several standard physical assumptions and requires abandoning at least one of the basic tenets of modern physics, we have to understand  correlations between ingoing and outgoing states of matter and gravity.

These results are consistent with the astrophysical observations. The effects of black hole radiation are negligible for the observed astrophysical black holes. Indeed, the classical event horizon is both the signature of black holes and the asymptotic benchmark against which the data available to the distant observers is compared~\cite{fn:98,af:13,ligo:16}. Change of the asymptotic limit by the  amount  of the order of $\epsilon_*$ that is constant on the cosmological time scales, will not affect the observations.

 {With a true Hamiltonian such as \eqref{Hred}, unitarity of the  combined matter-gravity system in a full quantum theory is immediate. The essential problem is how much entanglement between matter and gravity is needed to satisfy the constraint \eqref{consop}. The physical Hamiltonian in some time gauge would also create entanglement. Among the questions of interest is how the matter entropy, {obtained from a pure gravity-matter density {matrix} \cite{ht:10,kay:15}}, evolves at late times.}

 { In particular, it is important to understand  how the   intermode correlations of \cite{ss:15} behave when both the shell and the field are dynamical objects. While it is plausible that the constraint \eqref{consop} does not impose much matter-gravity entanglement both initially and in the the asymptotically flat  late spacetime, it will be instructive to understand if the gravitational degrees of freedom play a role of a catalyst (a subsystem whose state is unchanged in the overall evolution, but without which the transformation $\rho_\mathrm{in}\rightarrow\rho_\mathrm{out}$ is impossible, \cite{qinfo2,h4:09}).}

 { If the fully formed horizon does not exist, it is important to  investigate how (if at all) the soft hair properties of black holes \cite{gu:15,hps:16} are modified, as well whether the shell-radiation entanglement build-up follows the lines of the late information retrieval model \cite{bpz:13}.}

 There might be a lower bound on the lost information that is determined by a fundamental physics and not just by experimental or budgetary constraints. Indeed, the fundamental discreteness of space time, as well as quantumness of our measurement devices make   an ideal unitary evolution to appear as non-unitary \cite{gam}. In addition, the reasoning of \cite{mac} is applicable to black holes as well as to a quantum cosmological setting: the state may be pure, but still can appear to posses a non-zero entropy.

 Nevertheless,  seeking information recovery using Figure~\ref{figure1} as an axiom is flawed. The increase of entropy is not a sign of information loss, but a measure of redistribution of information between matter and gravitational degrees of freedom.

%\section{Conclusions}%if this part need
%\label{sec6}
%%%%%%%%%%%%%%%%%%%%%%%%%%%%%%%%%%%%%%%%%%

%%%%%%%%%%%%%%%%%%%%%%%%%%%%%%%%%%%%%%%%%%
\acknowledgments{We thank Mile Gu  and Robert Mann for collaboration on various issues that are discussed in this work, Jorma Louko for discussions, { Bernard Kay for useful suggestions}, Stefano Liberati and Bill Unruh for {suggestions and} critical comments.
 {Valentina Baccetti acknowledges support via the Macquarie Research Fellowship scheme.} }

%%%%%%%%%%%%%%%%%%%%%%%%%%%%%%%%%%%%%%%%%%
%% optional

%

%=====================================
% References, variant A: internal bibliography
%=====================================


\begin{thebibliography}{99}
\bibitem{h:74} Hawking, S.W. Black hole explosions? {\em Nature} \textbf{1974}, {\em 248}, 30--31.

\bibitem{h:76} Hawking, S.W. Breakdown of predictability in gravitational collapse. \emph{Phys. Rev. D} \textbf{1976}, \emph{14}, 2460--2473.

\bibitem{bmps:95} Brout, R.; Massar, S.; Parentani, R.; Spindel, P. A primer for black hole quantum physics. \emph{Phys. Rep.} \textbf{1995}, \emph{260}, 329--446.

\bibitem{fn:98} Frolov, V.; Novikov, I. \emph{Black Hole Physics: Basic Concepts and New Developments}; {{Kluwer}: Dordrecht, The~Netherlands,} 1998.
%please confirm the change of publisher: Kluwer-->Springer:: It was Kluwer in 1998

%\bibitem{wald:01lrr} Wald, R.M. \emph{Living. Rev. Relativ.} \textbf{2001}, \emph{4}, 6.
\bibitem{wald:01lrr} {Wald, R.M.} The Thermodynamics of Black Holes. In \emph{Advances in the Interplay Between Quantum and Gravity Physics}; Springer: Dordrecht, The~Netherlands, 2002; pp. 477--522. %please confirm the change of this reference
    %%% I'm not familiar with this reference.

\bibitem{modern} Mann, R.B. \emph{Black Holes: Thermodynamics, Information, and Firewalls}; Springer: New York, NY, USA, 2015.

\bibitem{modern2} {Harlow, D.} Jerusalem lectures on black holes and quantum information. \emph{Rev. Mod. Phys.} \textbf{2016}, \emph{88}, 015002.
%please confirm the change of name: Harlow, R.B.-->Harlow, D

\bibitem{qinfo} Bru{\ss}, D.; Leuchs G. (Eds.) \emph{Lectures on Quantum Information}; Wiley-VCH: Weinheim, Germany, 2007.

\bibitem{qinfo2} Wilde, M.W. \emph{Quantum Information Theory}; Cambridge University Press: Cambridge, UK, 2013.




\bibitem{rqi} Peres, A.; Terno, D.R. Quantum information and relativity theory. \emph{Rev. Mod. Phys.} \textbf{2004}, \emph{76}, 93.

\bibitem{rqi2} Mann, R.B.; Ralph, T.C. Relativistic quantum information. \emph{Class. Quant. Grav.} \textbf{2012}, \emph{29},(22). 

\bibitem{h4:09} Horodecki,R.; Horodecki, P., Horodecki, M., Horodecki, K. Quantum entanglement. \emph{Rev. Mod. Phys.} \textbf{2009}, \emph{81}, 865.

\bibitem{t:99} Terno, D.R. Nonlinear operations in quantum-information theory. \emph{Phys. Rev. A } \textbf{1999}, \emph{59}, 3320.

\bibitem{s:93} Susskind, L.; Thorlacius, L.; Uglum, J. The stretched horizon and black hole complementarity. \emph{Phys. Rev. D} \textbf{1993}, \emph{48}, 3743.

\bibitem{amps} Almhieri, A.; Marolf, D.; Polchinski, J.; Sully, J. Black Holes: Complementarity or Firewalls? \emph{JHEP}  \textbf{2013}, \emph{02}, 062.


\bibitem{bpz:13} Braunstein, S.L.; Pirandola, S.; \.{Z}yczkowski, K. Better Late than Never: Information Retrieval from Black Holes. \emph{Phys. Rev. Lett.} \textbf{2013}, \emph{110}, 101301.

\bibitem{kate} Bryan, K.L.H.; Medved, A.J.M. Black holes and information: A new take on an old paradox. {2016}, {arXiv:1603.07569}. 

\bibitem{mt:00} Mor, T.; Terno, D.R. Sufficient conditions for a disentanglement. \emph{Phys. Rev. A} \textbf{1999}, \emph{60}, 4341.

\bibitem{mile:15} Yuan, X.; Assad, S.M.; Thompson, J.; Haw, J.Y.; Vedral, V.; Ralph, T.C.;  Lam, P.K.; Weedbrook, C.; Gu, M. Replicating the benefits of Deutschian closed timelike curves without breaking causality. \emph{NPJ Quantum Inf.} \textbf{2015}, \emph{1}, 15007.%Please include the first ten authors’ names before using et al. in references.

\bibitem{jorma:14} Louko, J. Unruh-DeWitt detector response across a Rindler firewall is finite. \emph{JHEP} \textbf{2014}, \emph{1409}, 142.

\bibitem{ms:13} Maldacena, J.; Susskind, L. Cool horizons for entangled black holes. \emph{Fortschr. Phys.} \textbf{2013}, {\emph{61}}, 781--811.

\bibitem{topc} Friedman, J.L.; Schleich, K.; Witt D.M. Topological Censorship. \emph{Phys. Rev. Lett.} \textbf{1995}, \emph{75}, 1872.

\bibitem{misner} Misner, C.W. Wormhole Initial Conditions. \emph{Phys. Rev.} \textbf{1960}, \emph{118}, 1110.

\bibitem{local} Barat, N.; Kimball, J.C. Localization and causality for a free particle. \emph{Phys. Lett. A} \textbf{2003}, \emph{308}, 110--115.

\bibitem{local2} {Terno, D.R. Localization of relativistic particles and uncertainty relations. \emph{Phys. Rev. A} \textbf{2014}, \emph{89}, 042111.}
%ref 25  is not mentioned in mian text, please check and revise

\bibitem{hm:04} Horowitz G.T.; Maldacena, J.M. The black hole final state. \emph{JHEP} \textbf{2004}, \emph{02}, 008.

\bibitem{lp:14} Lloyd, S.; Preskill, J. Unitarity of black hole evaporation in final-state projection models. \emph{JHEP} \textbf{2014}, \emph{1408}, 126.

\bibitem{l:ctc} Lloyd, S.; Maccone, L.; Garcia-Patron, R.; Giovannetti, V.; Shikano, Y.; Pirandola, S.; Rozema, L.A.; Darabi, A.; Soudagar, Y.; Shalm, L.K.; et al. Closed Timelike Curves via Postselection: Theory and Experimental Test of Consistency. \emph{Phys. Rev. Lett.} \textbf{2011}, \emph{106}, 040403.
    
\bibitem{bv:14} Baez, J.C.; Vicary, J. Wormholes and entanglement. \emph{Class. Quant. Grav.} \textbf{2014}, \emph{31}, 

\bibitem{pw:00} Parikh, M.K.; Wilczek, F. Hawking Radiation As Tunneling. \emph{Phys. Rev. Lett.} \textbf{2000}, \emph{85}, 5042.

\bibitem{svzr:10} Singleton, D.; Vagenas, E.C.; Zhu, T.; Ren, J.R. Insights and possible resolution to the information loss paradox via the tunneling picture. \emph{JHEP} \textbf{2010}, \emph{1008}, 089;  erratum \emph{JHEP} \textbf{2011}, \emph{1101}, 021.

\bibitem{bp:11} Braunstein, S.L.; Patra, M.K. Black Hole Evaporation Rates without Spacetime. \emph{Phys. Rev. Lett.} \textbf{2011}, \emph{107}, 071302.

\bibitem{zczy:11} Zhang, B.; Cai, Q.Y.; Zhan, M.S.; You, L. Entropy is conserved in Hawking radiation as tunneling: A revisit of the black hole information loss paradox. \emph{Ann. Phys.} \textbf{2011}, \emph{326}, 350--363.

\bibitem{ijmpd:13} Zhang, B.; Cai, Q.Y.; Zhan, M.S.; You, L. Information conservation is fundamental: Recovering the lost information in hawking radiation. \emph{Int. J. Mod. Phys. D} \textbf{2013}, \emph{22}, 1341014.

\bibitem{ger:76} Gerlach, U.H. The mechanism of blackbody radiation from an incipient black hole. \emph{Phys. Rev. D} \textbf{1976}, \emph{14}, 1479.
%\bibitem{boulware:76} D.G. Boulware, Phys. Rev. D 13, 2169 (1976).

\bibitem{acvv:01} Alberghi, G.L.; Casadio, R.; Vacca, G.P.; Venturi, G. Gravitational collapse of a radiating shell. \emph{Phys. Rev. D} \textbf{2001}, \emph{64}, 104012.

\bibitem{liberati:06} Barcel\'{o}, C.; Liberati, S.; Sonego, S.; Visser, M. \emph{Class. Quantum Grav.} \textbf{2006}, \emph{23}, 5341.

\bibitem{liberati:08} Barcel\'{o}, C.; Liberati, S.; Sonego, S.; Visser, M. Fate of gravitational collapse in semiclassical gravity. \emph{Phys. Rev. D} \textbf{2008}, \emph{77}, 044032.

\bibitem{vsk:07} Vachaspati, T.; Stojkovic, D.; Kraus, L.M. Observation of incipient black holes and the information loss problem. \emph{Phys. Rev. D} \textbf{2007}, \emph{76}, 024005.

\bibitem{kmy:13} Kawai, H.; Matsuo, Y.; Yokokura, Y. A self-consistent model of the black hole evaporation. \emph{Int. J. Mod. Phys. A} \textbf{2013}, \emph{28}, 1350050.


\bibitem{poisson}Poisson, E. \emph{A Relativist's Toolkit}; Cambridge University Press: Cambridge, UK, 2004.

\bibitem{giddings:16} Giddings, S.B. Hawking radiation, the Stefan–Boltzmann law, and unitarization. \emph{Phys. Lett. B} \textbf{2016}, \emph{754}, 39--42.

\bibitem{dfu:76} Davies, P.C.W.; Fulling, S.A.; Unruh, W.G. Energy-momentum tensor near an evaporating black hole. \emph{Phys. Rev. D} \textbf{1976}, \emph{13}, 2720.

\bibitem{ss:15} Saini, A.; Stojkovic, D. Radiation from a Collapsing Object is Manifestly Unitary. \emph{Phys. Rev. Lett.} \textbf{2015}, \emph{114}, 111301.

\bibitem{vai:51} Vaidya, P.C. Nonstatic Solutions of Einstein's Field Equations for Spheres of Fluids Radiating Energy. \emph{Phys. Rev.} \textbf{1951}, \emph{83}, 10.

\bibitem{fw:99} Parikh, M.K.; Wilczek, F. Global structure of evaporating black holes. \emph{Phys. Lett. B} \textbf{1999}, \emph{449}, 24--29.

\bibitem{bmt:16} Baccetti, V.; Mann, R.B.; Terno, D.R. Role of evaporation in gravitational collapse. {2016}, {arXiv:1610.07839}. 

\bibitem{kay:98a} Kay, B.S. Entropy Defined, Entropy Increase and Decoherence Understood, and Some Black-Hole Puzzles Solved. {1998}, {arXiv:hep-th/9802172}. 

\bibitem{kay:98b} Kay, B.S. Decoherence of macroscopic closed systems within Newtonian quantum gravity. \emph{Class. Quant. Grav.} \textbf{1998}, \emph{15},L89-L98.

\bibitem{choptuik} Choptuik, M.W. Universality and scaling in gravitational collapse of a massless scalar field. \emph{Phys. Rev. Lett.} \textbf{1993}, \emph{70}, 9.

\bibitem{hw-flat} {Husain, V.}; Winkler, O. Flat slice Hamiltonian formalism for dynamical black holes. \emph{Phys. Rev. D} \textbf{2005}, \emph{71}, 104001.
\bibitem{ht:10} {Husain, V.}; Terno, D.R. Dynamics and entanglement in spherically symmetric quantum gravity. \emph{Phys. Rev. D }\textbf{2010}, \emph{81}, 044039.


\bibitem{ctz:08} Carteret, H.A.; Terno, D.R.; \.{Z}yczkowski, K. Dynamics beyond completely positive maps: Some properties and applications. \emph{Phys. Rev. A} \textbf{2008}, \emph{77}, 042113.

\bibitem{brustein:14}  R. Brustein, Origin of the blackhole information paradox. \emph{Fortschr. Phys.} \textbf{2014}, \emph{62}, 255-265. 

\bibitem{af:13} Abramowicz, M.A.; Fragile, P.C. Foundations of Black Hole Accretion Disk Theory. \emph{Living Rev. Relativ.} \textbf{2013}, \emph{16}, 1.

\bibitem{ligo:16} {LIGO Scientific Collaboration and Virgo Collaboration. Observation of Gravitational Waves from a Binary Black Hole Merger. \emph{Phys. Rev. Lett.} \textbf{2016}, \emph{116}, 061102.}

\bibitem{kay:15} Kay, B.S. Entropy and Quantum Gravity. \emph{Entropy} \textbf{2015}, \emph{17}, 8174--8186.

\bibitem{gu:15} G\"{u}rlebeck, N. No-Hair Theorem for Black Holes in Astrophysical Environments. \emph{Phys. Rev. Lett.} \textbf{2015}, \emph{114}, 151102.

\bibitem{hps:16} Hawking, S.W.; Perry, M.J.; Strominger, A. Soft Hair on Black Holes. \emph{Phys. Rev. Lett.} \textbf{2016} \emph{116}, 231301.

\bibitem{gam} Gambini, R.; Pullin, J. Relational Physics with Real Rods and Clocks and the Measurement Problem of Quantum Mechanics. \emph{Found. Phys.} \textbf{2007}, \emph{37}, 1074--1092.

\bibitem{mac} Maccone, L. Quantum Solution to the Arrow-of-Time Dilemma. \emph{Phys. Rev. Lett.} \textbf{2009}, \emph{103}, 080401.


\end{thebibliography}
\end{document}